\documentclass [12pt]{article}
\usepackage{amsfonts,amsmath,amssymb}
\usepackage{graphicx}
\usepackage{epic,eepic,color,graphicx}

\newfont{\twelvemsb}{msbm10 scaled\magstep1}
\newfont{\eightmsb}{msbm8}
\newfam\msbfam
\textfont\msbfam=\twelvemsb \scriptfont\msbfam=\eightmsb
\catcode`\@=11
\def\Bbb{\ifmmode\let\next\Bbb@\else
\def\next{\errmessage{Use \string\Bbb\space only in math mode}}\fi\next}
\def\Bbb@#1{{\fam\msbfam{{#1}}}}

\newcommand{\be}{\begin{equation}}
\newcommand{\ee}{\end{equation}}
\newcommand{\ba}{\begin{eqnarray}}
\newcommand{\ea}{\end{eqnarray}}

\begin{document}

\sloppy
\renewcommand{\thefootnote}{\fnsymbol{footnote}}
\newpage
\setcounter{page}{1} \vspace{0.7cm}
\vspace*{1cm}
\begin{center}
{\bf On the logarithmic powers of $sl(2)$ SYM$_4$ }\\
\vspace{1.8cm} {\large Davide Fioravanti $^a$, Paolo Grinza $^b$ and
Marco Rossi $^c$
\footnote{E-mail: fioravanti@bo.infn.it, pgrinza.grinza@usc.es, rossi@cs.infn.it}}\\
\vspace{.5cm} $^a$ {\em Sezione INFN di Bologna, Dipartimento di Fisica, Universit\`a di Bologna, \\
Via Irnerio 46, Bologna, Italy} \\
\vspace{.3cm} $^b${\em Departamento de Fisica de Particulas,
Universidad de Santiago de Compostela, 15782 Santiago de
Compostela, Spain} \\
\vspace{.3cm} $^c${\em Dipartimento di Fisica dell'Universit\`a
della Calabria and INFN, Gruppo collegato di Cosenza, I-87036
Arcavacata di Rende, Cosenza, Italy} \\
\end{center}
\renewcommand{\thefootnote}{\arabic{footnote}}
\setcounter{footnote}{0}
\begin{abstract}
{\noindent In the high spin limit the minimal
anomalous dimension of (fixed) twist operators in the $sl(2)$ sector of planar ${\cal N}=4$ Super Yang-Mills theory expands as $\gamma(g,s,L)=f(g) \ln s + f_{sl}(g,L) + \sum \limits _{n=1}^\infty \gamma^{(n)}(g,L)  \, (\ln s)^{-n} + \ldots $. We find that the sub-logarithmic contribution $\gamma^{(n)}(g,L) $ is governed by a linear integral equation, depending on the solution of the linear integral equations appearing at the steps $n'\leq n-3$. We work out this recursive procedure and determine explicitly $\gamma^{(n)}(g,L) $ (in particular $\gamma^{(1)}(g,L)=0$ and $\gamma^{(n)}(g,2)=\gamma^{(n)}(g,3)=0$). Furthermore, we  connect the $\gamma^{(n)}(g,L) $ (for finite $L$) to the generalised scaling functions, $f^{(r)}_n(g)$, appearing in the limit of large twist  $L\sim\ln s$. Finally, we provide the first orders of weak and strong coupling for
the first $\gamma^{(n)}(g,L)$ (and hence $f^{(r)}_n(g)$).}
\end{abstract}
\vspace{5cm}
{\noindent {\it Keywords}}: Integrability; Infinite Conserved Charges; Bethe Ansatz equations; AdS- CFT correspondence. \\

\newpage

\section{A quick outlook}
\setcounter{equation}{0}

The calculation of anomalous dimensions in the planar ${\cal N}=4$ Super Yang-Mills (SYM) theory received a considerable boost as a consequence of the hints for an underlying integrable structure (cf. for instance \cite{L-MZ, BS,AFS,ES,BES}). This meant the identification of the Bethe(-Yang) Ansatz equations valid in the asymptotic regime of very long operators. The energies (i.e. the integrable asymptotic spectrum) corresponds to planar anomalous dimensions of single trace operators, when the number of the compounds is very large. This approximation would correspond to the on-shell (or IR) description in the {\it usual} relativistic field theories \cite{ZZ}. Interestingly, in this regime it was possible to predict, in a relatively simple fashion, some exact weak-coupling expansions of anomalous dimensions up to a certain number of loops; these were successfully compared with results of field theoretical perturbation theory (e.g. \cite{BES, KL, KLRSV} and references therein). In some lucky cases, this approximation gave also the impressive access to some exact results in the strong coupling regime, dominated by the dual string theory (cf. e.g. \cite{GKPII-FT} and references therein). More in general, an anomalous dimension is affected by corrections coming from the finite size of a composite operator, but this on-shell approach has also furnished a basis for the off-shell (any length at any coupling) prediction of the recent Thermodynamic Bethe Ansatz \cite{TBA}. All these integrability based results are precision tests for the more general AdS/CFT correspondence \cite{MWGKP}.

In this paper we focus on the study of the twist $sl(2)$ sector of ${\cal N}=4$ planar SYM whose operators have the form
\ba
\textrm{Tr}({\mathcal D}^s {\mathcal Z}^L) + \dots \, , \label {twop}
\ea
where the $s$ symmetrised, traceless, covariant derivatives $\mathcal D$ act in all the possible ways on the $L$ bosonic fields ${\mathcal Z}$. $s$ is just the Lorentz spin and $L$ an $su(4)$ R-charge equal to the twist ($=$(classical dimension) - $s$). Peculiarly, the minimal anomalous dimension of (\ref {twop}) shows a leading high spin (fixed $L$) behaviour with the logarithm of $s$, having as coefficient the so-called {\it universal scaling function}, $f(g)$\footnote{It does not depend on $L$ and equals twice the {\it cusp anomalous dimension} of light-like Wilson loop.}  \cite{KM,BGK, FTT}.  Impressively, it was obtained from the solution of a linear integral equation directly derived from the asymptotic Bethe Ansatz (ABA) via the root density approach \cite {BES}. Moreover, it was carefully studied and tested both in the weak \cite {ES,BES} and strong coupling limit \cite {BBKS,CK,BKK,KSV}.

In the following, we wish to prove that the high spin expansion goes on as a series of logarithmic (inverse)
powers\footnote {With $O\left ( (\ln s )^{-\infty} \right )$ we indicate terms going to zero faster than any inverse powers of $\ln s $.},
\ba
\gamma(g,s,L)= f(g) \ln s + f_{sl}(g,L) + \sum_{n=1}^\infty \gamma^{(n)}(g,L)  \, (\ln s)^{-n} + O\left ( (\ln s )^{-\infty} \right ) \, , \label {subl}
\ea
i.e. it is a genuine large size expansion in the size parameter $\ln s$. In this context
the sub-leading (constant) contribution $f_{sl}(g,L)$ received already much attention. In \cite
{FRS} it was shown to come from the solution of a non-linear integral equation (NLIE). Then, in \cite{BFR} it was obtained starting from a linear integral equation (LIE). Explicit weak and strong coupling expansions are present in \cite {FZ} and agree with string theory computations \cite {BFTT}. Importantly, it is believed that both $f(g)$ and $f_{sl}(g,L)$ are exactly given by this approach based on the ABA without finite size corrections. In fact, for instance the findings of
\cite {BJL} have proved that, at least at twist two and up to four loops, wrapping corrections
start contributing at order $O\left ((\ln s )^2 /s \right )$. More in general, we may expect that all the logarithmic sub-leading contributions in (\ref {subl}), $\gamma^{(n)}(g,L)$, are exact as they can be elaborated from the ABA.

In this letter we focus on $\gamma^{(n)}(g,L)$. We prove that each $\gamma^{(n)}(g,L)$ is still determined by a {\it linear} integral equation (for the density of roots and holes), which can be equivalently rewritten as an infinite dimensional (matrix) linear system. Things can be arranged in such a way that for any $n$ the systems have the same kernel\footnote {This is indeed the BES kernel determining $f(g)$ \cite{BES} and $f_{sl}(g,L)$ \cite{BFR, FZ}.}, differing only in the inhomogeneous terms. These forcing terms are expressed as linear combinations of those
driving the systems for the ``reduced coefficients", defined by equations (4.23,24) of \cite {FGR3} and appearing in the study of the so-called generalised scaling functions $f_m(g)$  \cite {FRS, FGR3} and $f_m^{(r)}(g)$, describing the anomalous dimension
\be
\gamma (g,s,L)=\ln s \sum ^{\infty}_{n=0} f_n(g) j^n + \sum _{r
=0}^{\infty} (\ln s )^{-r}\sum
^{\infty}_{n=0}
f_n^{(r)}(g) j^n + O\left ( (\ln s )^{-\infty} \right ) \, , \label {gammaj}
\ee
in the limit \cite {BGK}
\be
s\rightarrow \infty \, , \quad L\rightarrow \infty \, , \quad j=\frac {L-2}{\ln s}  \, \quad {\mbox {fixed}} \, .
\label {jlimit}
\ee
In this way a recursive structure is set up such that the driving term in the system $n$ for $\gamma^{(n)}(g,L)$ involves the solutions to those with $n' \leq n-3$. Therefore, it is possible to push the computations at any desired order $n$ and give a general expression for the form of $\gamma^{(n)}(g,L)$. From the latter the higher generalised scaling functions $f_m^{(r)}(g)$ can be extracted \footnote{They can also be computed independently as in Appendix B}.

\section{All-loops ABA and the (N)LIE at high spin}
\setcounter{equation}{0}

In a series of papers we proposed the NLIE \cite {NOI} technique for the study of asymptotic Bethe Ansatz equations describing the AdS/CFT correspondence.
An interesting case where such a technique was recently applied \cite{BFR} is the calculation of anomalous dimensions of single-trace operators belonging to the $sl(2)$ sector of the ${\cal N}=4$ SYM theory.
In such a specific case, if one looks at the high spin limit at fixed twist, one has two crucial simplifications. Firstly, the non-linear integral contributions to the NLIE are depressed and contribute with terms  $O\left ( (\ln s)^{-\infty} \right )$ \cite{FRS,BFR}. Secondly -  as stated in the introduction - wrapping effects are also depressed and start contributing at order
$O((\ln  s)^2/s)$ \cite{BJL}. This means that in order to work out all the sub-leading terms behaving as $(\ln s)^{-n}$, with $n \ge 0$, we can safely rely on the linear integral terms of the NLIE coming from the asymptotic Bethe Ansatz.

In this letter we focus on the minimal anomalous dimension state, with spin $s$ and twist $L$.
Such a state is characterised by $s$ Bethe roots, localised in an interval $[-b, b]$ of the real axis
and by $L$ real 'holes'. The positions of both the roots and the holes are symmetric with respect to the origin.
For what concerns the holes, two lie outside $[-b, b]$, the remaining $L-2$ concentrate near the origin, with no roots lying in between.
The positions $\upsilon _k$ of both roots and holes satisfy the condition $Z(\upsilon _k)=\pi (2k+1)$, with $k$ integer,
where $Z(u)=-Z(-u)$ is the counting function; the positions of the internal holes $u_h$ are unknowns in the NLIE and are determined by the (non linear) relations $Z(u_h)=\pi (2h+1-L) \, , \quad h=1, \ldots , L-2 \, $, which supplement the (linear integral part of the) NLIE.

We move from upshots of \cite{BFR}. The crucial point is that the forcing term $F(u)$ defined in \cite{BFR} approximates the counting function $Z(u)$ in the large $s$ limit, if we neglect terms of order $O\left ( (\ln s)^{-\infty} \right )$.
Hence, the density of roots and holes $\sigma(u) = \frac{d}{du}Z(u)$ can be computed (up to this order) by using the linear integral equation (3.52, 4.10 of \cite{BFR}) for $F(u)$.
In specific, splitting the forcing term in one and higher than one loops contributions, $F(u) = F_0(u) + F^H(u)$, the linear integral equation for the higher-loop part $F^H(u)$ is\footnote {For notations we refer to seminal papers \cite {BS,AFS,ES,BES}.}
\ba
F^H(u)&=&-iL \ln \left ( \frac {1+\frac {g^2}{2x^-(u)^2}}{1+\frac {g^2}{2x^+(u)^2}}\right ) -2i \sum _{h=1}^{L-2}
\Bigl [ \ln \left ( \frac {1-\frac {g^2}{2x^+(u)x^-(u_h)}}{1-\frac {g^2}{2x^-(u)x^+(u_h)}}\right ) + \nonumber \\
&+& i \theta (u,u_h)+ i \arctan (u-u_h)-i \arctan (u-u_h^{(0)}) \Bigr ]+ \int _{-\infty}^{+\infty} \frac {dv}{\pi}
\frac {1}{1+(u-v)^2}F^H(v) + \label {FHu}  \\
&+& \frac {i}{\pi} \int _{-\infty}^{+\infty} dv \left [ \frac {d}{dv}\ln \left ( \frac {1-\frac {g^2}{2x^+(u)x^-(v)}}{1-\frac {g^2}{2x^-(u)x^+(v)}}\right )+ i \frac {d}{dv} \theta (u,v) \right ] [F_0(v)+F^H(v)] +
O\left ( (\ln s)^{-\infty} \right ) \, . \nonumber
\ea
This equation comes from (4.10) of \cite{BFR}, after removing all the $O\left ( (\ln s)^{-\infty} \right )$ terms.
It has to be supplemented with the expression for $F_0(u)$ - (3.52) of \cite {BFR} - which in Fourier transform
reads:
\be
ik \hat F_0(k)=-4\pi \frac {\frac {L}{2}-e^{-\frac {|k|}{2}}\cos (ks /{\sqrt 2}) }{2 \sinh \frac {|k|}{2}}
+2\pi \sum _{h=1}^{L-2}e^{iku_h^{(0)}} \frac {e^{-\frac {|k|}{2}}}{2 \sinh \frac {|k|}{2}} -4\pi \ln2 \delta (k)
+ O\left ( (\ln s)^{-\infty} \right ) \, . \label {F0u}
\ee
In (\ref {FHu}, \ref {F0u}) $u_h^{(0)}$ represent the one loop contribution to the $g$-depending position of the internal holes $u_h$.
Now, since the function $F(u)$ approximates the counting function $Z(u)$ up to $O\left ( (\ln s)^{-\infty} \right )$ terms, in this approximation the positions of the internal holes $u_h$ are determined from the conditions
\be
F(u_h)=\pi (2h+1-L) \, , \quad h=1, \ldots , L-2 \, .
\label{condhole}
\ee
The key point which allows to understand the origin of the logarithmic terms $(\ln s)^{-n}$ is related to the behaviour of the position of the holes as a function of the spin $s$ in the $s \to \infty$ limit. In such a limit for fixed $u$ $F(u)$ diverges logarithmically   (and expands in (inverse) powers of $\ln s $)
as
\be
F(u)=\sum _{n=-1}^\infty F^{(n)}(u) (\ln s)^{-n} + O\left ( (\ln s)^{-\infty} \right ) \, ,
\ee
hence it is natural to suppose that - in order to fulfil (\ref {condhole}) - the position of the holes
has to expand in inverse powers of $\ln s$:
\ba
u_h = \sum_{n=1}^\infty \alpha_{n,h} (\ln s)^{-n} + O\left ( (\ln s)^{-\infty} \right ) \, . \label {upos}
\ea
A systematic way to perform such an expansion order by order in a recursive way is given by the Fa\`a di Bruno formula for the derivatives of a composite function.
We already used this trick in \cite{FGR3} and we refer to that paper for technical details.
Introducing the derivatives in zero of the function $\sigma (u)=\sigma (-u)=\frac{d}{du}F(u)$ and developing them in powers of $\ln s$,
\be
\frac{d^{r}}{d u^{r}} \sigma(u=0) = \sum_{n=-1}^\infty \sigma^{(n)}_{r} \, (\ln s)^{-n} \, , \label {deriv}
\ee
($ \sigma^{(n)}_{r}=0$ when $r$ is odd) the condition (\ref {condhole}) for the holes eventually gives
\ba
\pi (2h+1-L) &  =  & \sigma^{(-1)}_0 \alpha_{1,h} + \sum_{p=1}^{\infty} (\ln s)^{-p}
\, \sum_{r=1}^{p+1}  \sigma^{(-1)}_{r-1} \sum_{\{ j_1, \dots, j_{p-r+2}\}}
\prod_{m=1}^{p-r+2} \frac{(\alpha_{m,h})^{j_m}}{j_m !} + \nonumber \\
& + & \sum_{p=1}^{\infty} (\ln s)^{-p}
\, \sum_{l=0}^{p-1}   \sum_{r=1}^{p-l}   \sigma^{(l)}_{r-1} \sum_{\{ j_1, \dots, j_{p-r-l+1}\}}
\prod_{m=1}^{p-r-l+1} \frac{(\alpha_{m,h})^{j_m}}{j_m !} \, , \label {condhole2}
\ea
where the $j_m$ contained in the second term of the r.h.s. are constrained by the conditions
$\sum _{m=1}^{p-r+2}j_m=r$, $\sum _{m=1}^{p-r+2}mj_m=p+1$, the ones in the third term
by $\sum _{m=1}^{p-r-l+1}j_m=r$, $\sum _{m=1}^{p-r-l+1}mj_m=p-l$.
Equating l.h.s. and r.h.s. at all orders in $\ln s$ we obtain the following recursive equation
\ba
\alpha_{p+1,h}  &  =  & -  \sum_{r=1}^{p} \frac{\sigma^{(-1)}_{r}}{ \sigma^{(-1)}_0} \sum_{\{ j_1, \dots, j_{p-r+1}\}}
\prod_{m=1}^{p-r+1} \frac{(\alpha_{m,h})^{j_m}}{j_m !} -\sum_{l=0}^{p-1}   \sum_{r=1}^{p-l} \frac{  \sigma^{(l)}_{r-1}}{ \sigma^{(-1)}_0} \sum_{\{ j_1, \dots, j_{p-r-l+1}\}}
\prod_{m=1}^{p-r-l+1} \frac{(\alpha_{m,h})^{j_m}}{j_m !}, \ \ \ p\geq 1 \nonumber \\
\alpha_{1,h}  &  =  & \frac{\pi (2h-1+L)}{\sigma^{(-1)}_0} \, ,
\label{alphaiter}
\ea
where now the $j_m$ contained in the first term of the r.h.s. are constrained by the conditions
$\sum _{m=1}^{p-r+1}j_m=r+1$, $\sum _{m=1}^{p-r+1}mj_m=p+1$ and the ones in the second term
by $\sum _{m=1}^{p-r-l+1}j_m=r$, $\sum _{m=1}^{p-r-l+1}mj_m=p-l$.
Equation (\ref {alphaiter}) is solved by iterations, allowing to express the coefficients $\alpha_{n,h}$ in terms of the derivatives in zero of the density of roots and holes. The first three of them are
\ba
\alpha_{1,h}&=& \frac{\pi (2h+1-L)}{  \sigma^{(-1)}_{0} } \, , \quad
\alpha_{2,h}= -\frac{\pi (2h+1-L)\sigma^{(0)}_{0}}{  (\sigma^{(-1)}_{0})^2 } \, , \quad  \nonumber \\
\alpha_{3,h}&=& \frac{\pi (2h+1-L)}{  \sigma^{(-1)}_{0} } \left( - \frac{\sigma^{(1)}_{0} }{  \sigma^{(-1)}_{0}} +  \left( \frac{\sigma^{(0)}_{0} }{  \sigma^{(-1)}_{0} } \right)^2
- \frac  {\pi^2}{6} (2h+1-L)^2 \frac{\sigma^{(-1)}_{2} }{ ( \sigma^{(-1)}_{0})^3}
\right) \, . \label {alfa}
\ea

\section{Linear integral equations for the logarithmic terms}
\setcounter{equation}{0}

After we clarified the behaviour of the position of the holes for high $s$, we continue with the standard treatment of equation (\ref{FHu}), in order to study the contributions to the anomalous dimension proportional to $(\ln s )^{-n}$.
For simplicity's sake, from now on we do not write the $O\left ( (\ln s)^{-\infty} \right )$ terms.  

Since the anomalous dimension at high spin is given
\cite {KL} by $\gamma(g,s,L) =  \lim\limits _{k \to 0} \frac{\hat \sigma _H(k)}{\pi}$,
where $\sigma _H(u)=\frac{d}{du} F^H(u)$, we write (\ref {FHu}) in terms of
the Fourier transform $\hat \sigma _H(k)$, restrict to $k>0$ and introduce
\ba
S(k)&=&\frac {\sinh \frac {|k|}{2}}{\pi |k|} \Bigl \{ \hat \sigma _H(k)
- \pi  \frac {e^{-\frac {|k|}{2}}}{ \sinh \frac {|k|}{2}}  \sum _{h=1}^{L-2} \left[ \cos k u_h-\cos { k u_h^{(0)}} \right]
\Bigr \} \, \Rightarrow \,\gamma(g,s,L) = 2 \lim_{k \to 0} S(k)  \, . \label{Sdef}
\ea
The function (\ref {Sdef}) satisfies the linear integral equation
\begin{eqnarray}
&&S(k)=\frac {L}{k}[1-J_0({\sqrt {2}}gk)]
- g^2 \int _{0 }^{+\infty}\frac {dt}{\pi }
e^{-\frac{t}{2}} \hat K (\sqrt{2}gk,\sqrt{2}gt) \cdot \nonumber \\
 &\cdot& \Bigl \{ \frac {\pi t}{\sinh \frac {t}{2}}S(t)-4\pi \ln 2 \ \delta (t)-\pi (L-2)
\frac {1-e^{\frac {t}{2}}}{\sinh \frac {t}{2}}-2\pi \frac {1-
e^{-\frac {t}{2}}\cos \frac {ts}{{\sqrt {2}}}}{\sinh \frac {t}{2}}
+ \label {Skeq2} \\
&+& \pi  \frac {e^{\frac {t}{2}}}{ \sinh \frac {t}{2}}  \sum _{h=1}^{L-2} \left[ \cos t u_h -1 \right] \nonumber
\Bigl \} = 4g^2 \ln s \ \hat K( \sqrt{2}gk, 0) + 4g^2 \int _{0 }^{+\infty} \frac{dt}{e^{t}-1} \hat K^{*} (\sqrt{2}gk,\sqrt{2}gt)+ \nonumber  \\
&+&  \frac {L}{k}[1-J_0({\sqrt {2}}gk)]+4g^2 \gamma _E \, \hat K( \sqrt{2}gk, 0) + g^2 (L-2) \int _{0 }^{+\infty}dt  e^{-\frac{t}{2}}\hat K (\sqrt{2}gk,\sqrt{2}gt)  \frac {1-e^{\frac {t}{2}}}{\sinh \frac {t}{2}}  - \nonumber \\
&-& g^2 \int _{0 }^{+\infty} {dt}
\hat K (\sqrt{2}gk,\sqrt{2}gt)
 \frac {\sum _{h=1}^{L-2} \left[ \cos t u_h -1 \right]}{ \sinh \frac {t}{2}}
 -
g^2 \int _{0 }^{+\infty}{dt}
e^{-\frac{t}{2}} \hat K (\sqrt{2}gk,\sqrt{2}gt)
  \frac { t}{\sinh \frac {t}{2}}S(t)
\, , \nonumber
\end{eqnarray}
where the 'magic' separable kernel $\hat K(t,t')$ is defined in \cite {BES} as
\be
\hat K(t,t')=\frac{2}{tt'}\left [ \sum _{n=1}^{\infty}n J_n (t) J_n (t') + 2 \sum _{k=1}^{\infty} \sum _{l=0}^{\infty} (-1) ^{k+l}c_{2k+1,2l+2}(g) J_{2k}(t) J_{2l+1}(t') \right ] \label {magic}
\ee
and its modification $\hat K^{*} (t,t')$ as (\ref {magic}) with the only replacement $J_1(t')\rightarrow
J_1(t')-t'/2$.
In writing (\ref {Skeq2}) the various contributions to the forcing term are separated according to their power of $\ln s$. The term proportional to $\ln s $ enters the BES equation for the cusp anomalous dimension, the four subsequent terms - independent of $s$ - appear in the equation for the density which determines the virtual scaling function \cite {FZ}. On the other hand,
the contributions proportional to $(\ln s )^{-n}$, $n\geq 1$, come from the term containing
\be
P(s,g,t)=\sum _{h=1}^{L-2} \left[ \cos t u_h -1 \right] \, . \label {Ps}
\ee
Expanding (\ref {Ps}) for high $s$ we have,
because of (\ref {upos}),
\be
P(s,g,t)  =   \sum_{n=1}^\infty \mathcal P_n (g,t) (\ln s)^{-n} \, .
\label{logexpan}
\ee
Using again the Fa\`a di Bruno formula, we can give a reasonably explicit expression for $\mathcal P_n (g,t)$. After some calculations we obtain
\ba
\mathcal P_n (g,t) =
\sum_{r=1}^n  t^r \; \cos \frac{\pi r}{2}\sum_{\{ j_1, \dots, j_{n-r+1}\}}
\frac{ \sum\limits_{h=1}^{L-2}\prod_{m=1}^{n-r+1} (\alpha_{m,h})^{j_m}}{\prod_{m=1}^{n-r+1}  j_m !}, \ \ \ \ \
\sum _{m=1}^{n-r+1}j_m=r \, , \ \ \ \sum _{m=1}^{n-r+1}m j_m=n \, .
\ea
{\bf Remark 1} From the structure of the various $ \alpha_{n,h}$ (see e.g. (\ref {alfa}) for $n=1,2,3$) it follows that $P(s,g,t)$ is zero if $\mathcal{S}(m,L)\equiv \sum \limits _{h=1}^{L-2} (2h+1-L)^m=0$, $\forall m \in \mathbb{N}$. This happens for $L=2,3$. Thus, we conclude that for $L=2,3$ the high spin expansion of the anomalous dimension does not contain inverse logarithmic powers in the spin, i.e. $\gamma^{(n)}(g,2)=\gamma^{(n)}(g,3)=0$, $\forall n \geq 1$.

\medskip

\noindent {\bf Remark 2} If we extend formally the result for $\mathcal{S}(m,L)$ to the case $L=1$, we find again $\mathcal{S}(m,L=1)=0$.
We will use this formal property later, when discussing the relation to the large twist limit.

\medskip

It is now natural to introduce the high $s$ expansion for the function $S(k)$,
\ba
S(k) = \sum_{n=-1}^{\infty} S^{(n)}(k) \, (\ln s)^{-n} \, ,
\ea
which allows to obtain a linear integral equation at each order in $(\ln s )^{-n}$. Let us focus on
$n \geq 1$. We immediately remark that $S^{(1)}(k) = 0$, because the $1/\ln s$ term is absent in the large $s$ expansion of $P(s,g,t)$.
The next step is the Neumann expansion for $S^{(n)}(k)$, which is a standard procedure \cite {KL} in the case of an integral equation with separable kernel,
 \be
S^{(n)}(k)=\sum _{p=1}^{\infty} S_p^{(n)}(g) \frac {J_p({\sqrt 2}gk)}{k} \, \Rightarrow
\gamma^{(n)}(g,L) = {\sqrt 2} g S_1^{(n)}(g) \, .
\ee
The Neumann expansion transforms the linear integral equation for $S^{(n)}(k)$ into a linear infinite
system\footnote {We use the notation:
\begin{equation}
Z_{n,m}(g)= \int _{0}^{+\infty}
\frac {dt}{t} \frac {J_{n}({\sqrt {2}}gt)J_{m}({\sqrt {2}}gt)}{e^t-1}
\, .
\end{equation}},
\begin{eqnarray}
S^{(n)}_{2p-1}(g)&=& -(2p-1) \int _{0}^{+\infty}
\frac{dt}{t} \frac {{\mathcal P}_n(g,t) \, J_{2p-1}({\sqrt {2}}gt)}{\sinh \frac {t}{2}} - 2(2p-1)
\sum _{m=1}^{\infty} Z_{2p-1,m}(g) S^{(n)}_m (g) \, , \nonumber \\
S^{(n)}_{2p}(g)&=& - 2p \int _{0}^{+\infty}
\frac{dt}{t} \frac { {\mathcal P}_n(g,t) \,   J_{2p}({\sqrt {2}}gt)}{\sinh \frac {t}{2}} -
4p\sum _{m=1}^{\infty} Z_{2p,m}(g) (-1)^m S^{(n)}_{m}(g) \, , \label {Ssystem}
\end{eqnarray}
for each of the Neumann modes $S_p^{(n)}(g)$. A look at (\ref {Ssystem}) shows that
$S_p^{(n)}(g)$ are expressed in terms of the ``reduced coefficients" $\tilde S^{(n)}_{p}(g)$ -
defined in (4.23, 4.24) of \cite{FGR3} - and the quantities (\ref {deriv}) $\sigma _{2q}^{(n')}$, with $n'\leq n-3$: this makes it possible to build up a recursive calculation scheme, opening the way to push the computation up to the desired order in $\ln s$, in a way similar to \cite{FGR3}. And, indeed, from (\ref {Ssystem}) we can get the expression for
$\gamma^{(n)} (g)$ in terms of $\tilde S^{(n)}_{1}(g)$ and $\alpha_{n,h}$ as
\ba
\frac{\gamma^{(n)} (g)}{\sqrt2 \, g} = -2 \pi \sum_{r=1}^n  \tilde S^{(r/2)}_{1}(g) \; \cos \frac{\pi r}{2}  \sum_{\{ j_1, \dots, j_{n-r+1}\}}
\frac{ \sum\limits_{h=1}^{L-2}\prod\limits _{m=1}^{n-r+1} (\alpha_{m,h})^{j_m}}{\prod\limits_{m=1}^{n-r+1}  j_m !} \, , \
\sum _{m=1}^{n-r+1}j_m=r \, , \,   \sum _{m=1}^{n-r+1}m j_m=n \, , \label {gammamain}
\ea
where $\alpha_{m,h}$ in (\ref{gammamain}) depend on $\sigma _{2q}^{(n')}$, with $n'\leq n-3$. Explicitly,
we have $\gamma^{(1)} (g,L)=0$ and
\ba
\gamma^{(2)} (g,L) & = & \sqrt 2 g \,  \frac{\pi^3}{3 \, (\sigma^{(-1)}_{0})^2} (L-3)(L-2)(L-1) \,  \tilde S^{(1)}_{1}(g) \, ,
\label {gamma2}\\
\gamma^{(3)} (g,L) & = & - 2 \sqrt 2 g \frac{\pi^3 \, \sigma^{(0)}_{0}}{3 \, (\sigma^{(-1)}_{0})^3} (L-3)(L-2)(L-1)  \, \tilde S^{(1)}_{1}(g) \, ,
\label {gamma3} \\
\gamma^{(4)} (g,L) & = & \sqrt 2 g \,  2 \pi  \Bigl \{ \Bigl [
 -\frac{\pi^2}{3 \, (\sigma^{(-1)}_{0})^2} \Bigl (  \frac{\sigma^{(1)}_{0} }{  \sigma^{(-1)}_{0}} -
 \frac{3}{2} \frac{(\sigma^{(0)}_{0})^2 }{  (\sigma^{(-1)}_{0})^2 }  \Bigl ) (L-3)(L-2)(L-1) -  \nonumber \\
&-&  \frac{\pi^4 \sigma^{(-1)}_{2} }{90 \, (\sigma^{(-1)}_{0})^5} (L-3)(L-2)(L-1)(5+3L(L-4))
 \Bigr ] \tilde S^{(1)}_1 (g) \,  -  \nonumber \\
 &-& \frac{\pi^4}{360 \, (\sigma^{(-1)}_{0})^4} (L-3)(L-2)(L-1)(5+3L(L-4)) \, \tilde S^{(2)}_1 (g) \Bigr \} \label {gamma4} \, .
\ea
These expressions are valid $\forall g$ and will be disentangled in the weak and strong coupling limit.
Weak coupling expansions are provided in Appendix A, the strong coupling leading
term in Section 5.

\section{Relation to the large twist case}
\setcounter{equation}{0}

The expressions for $\gamma^{(n)} (g,L)$ can be written in terms of the 'generalised' scaling functions $f_n(g)$, $f_n^{(r)}(g)$, describing - according to formula (\ref {gammaj}) - the anomalous dimension in the limit (\ref {jlimit}).

Beside the widely investigated $f_n(g)$, the functions $f_n^{(0)}(g)$ were studied in \cite {FIR}.
Expressions for $f_n^{(r)}(g)$, $r\geq 1$, can be obtained following the lines of Appendix A of \cite {FIR}
and are explicitly given, when $1\leq r \leq 4$ and $0\leq n \leq 4-r$, in Appendix B of this letter.
In the limit $L\rightarrow \infty$ expansion (\ref {subl}) coincides\footnote {This is true because we include all the terms $(\ln s )^{-n}$ in both expansions (the next
term would be of order $O(\ln s/s)$).} with (\ref {gammaj}), after using the definition $j=(L-2)/\ln s$.  Therefore, from the comparison between (\ref {subl}) and (\ref {gammaj}), one has, at large $L$,
\be
\gamma^{(n)} (g,L)= (L-2)^{n+1}f_{n+1}(g) + \sum _{r=0}^{n} (L-2)^{n-r}f_{n-r}^{(r)}(g) \, ,
\quad n \geq 1 \, .
\label {general}
\ee
Since $\gamma ^{(n)}(g,L)$ is a polynomial in $L-2$, (\ref {general}) can be extended to arbitrary $L$.
Now, we use the fact that $\gamma ^{(n)} (g,2)=\gamma ^{(n)}(g,3)=0$, (see Remark 1): this
implies, when $n\geq 1$, $f_0^{(n)}(g)=0$ and $f_{n+1}(g)+\sum \limits _{r=0}^{n}f_{n-r}^{(r)}(g)=0$, respectively.
In addition, the 'formal' property announced in Remark 2, $\gamma ^{(n)} (g,1)=0$, produces the other relation
$(-1)^{n+1}f_{n+1}(g)+\sum \limits _{r=0}^{n}(-1)^{n-r}f_{n-r}^{(r)}(g)=0$, $n\geq 1$.
Imposing such relations on the general expression (\ref {general}), one gets $\gamma^{(1)} (g,L)=0$ and, $\forall n \geq 2$,
\be
\gamma^{(n)} (g,L)=\left [(L-2)^{n+1}-(L-2)^{\frac{3+(-1)^{n+1}}{2}}\right ]f_{n+1}(g)+\sum _{r=0}^{n-3}\left [ (L-2)^{n-r} - (L-2)^{\frac{3+(-1)^{n-r}}{2}} \right ] f_{n-r}^{(r)}(g) \, ,
\label {gammagen}
\ee
which expresses $\gamma^{(n)} (g,L)$ in terms of the generalised scaling functions\footnote {The 'inverse' relation, which expresses the generalised scaling functions
in terms of $\gamma^{(n)} (g,L)$, comes from the derivatives of the latter with respect to $L$:
\be
f_m^{(n-m)}(g) = \frac{1}{m!}\frac{d^m}{dL^m}\gamma^{(n)} (g,L)|_{L=2} \, , \quad n \geq 1 \, , \quad 0 \leq m \leq n+1 \, .
\ee} at fixed $j$.

Let us now show that (\ref {gammagen}) agrees with formul{\ae} (\ref {gamma2} - \ref {gamma4}).
We first notice that $\sigma _0^{(1)}$, entering the expression of $\gamma ^{(4)}(g,L)$, is zero, since the density at the order $(\ln s )^{-1}$ vanishes. Then, using notations of \cite {FIR} (cf. formul{\ae} (4.13,15) of that paper), we remark
that $\sigma _0^{(0)}=(L-2) \sigma ^{(-1,1)}+ \sigma ^{(0,0)}$. Plugging such expression into
(\ref {gamma2} - \ref {gamma4}) and using results from \cite {FIR} and Appendix B (formula (\ref {f13})), we obtain
\ba
\gamma^{(2)} (g,L) & = &  [(L-2)^3-(L-2)] \, f_3(g)  \label {gamma2bis} \, ,  \\
\gamma^{(3)} (g,L) &= & [(L-2)^4-(L-2)^2] f_4(g)+[(L-2)^3-(L-2)]f_3^{(0)}(g) \label {gamma3bis} \, , \\
\gamma^{(4)} (g,L)&=&[(L-2)^5-(L-2)]f_5(g)+[(L-2)^4-(L-2)^2]f_4^{(0)}(g) + \label  {gamma4bis}\\
&+& [(L-2)^3-(L-2)]f_3^{(1)}(g) \nonumber \, ,
\ea
which indeed agree with (\ref {gammagen}).

\section{Strong coupling limit}
\setcounter{equation}{0}

We are now in the position to extract the leading strong-coupling behaviour of  (\ref {gamma2}-\ref {gamma4}) in an analytic way. To this purpose we briefly recall some results obtained in \cite{FGR3}.
At strong coupling, the first component of the solution of the ``reduced system" can be written as
\ba
\sqrt2 g \tilde S_1^{(n)} (g) =\frac{ (-1)^{n+1} }{2 \pi} \left( \frac{\pi}{2} \right)^{2n} m(g)+
O\left ( e^{-\frac{3\pi g}{\sqrt 2} } \right )
, \ \ \ \ \  n \geq 1
\ea
and, for the BES related densities $\sigma_{2q}^{(-1)} $, we get
\ba
\sigma_{2q}^{(-1)} = -\pi  \left( \frac{\pi}{2} \right)^{2q} m(g)+ O\left ( e^{-\frac{3\pi g}{\sqrt 2} } \right ) , \ \ \ \ \  q \ge 0 \, ,
\ea
where $m(g)$ is the mass gap of the $O(6)$ Non-Linear Sigma Model embedded \cite {BK} in ${\cal N}=4$ SYM
\ba
m(g) = \frac{2^{5/8} \pi^{1/4}}{\Gamma(5/4)} \, g^{1/4} \left[ 1+ O \left ( g^{-1} \right ) \right] \, e^{-\frac{\pi g}{\sqrt 2} } + \ldots \quad .
 \ea
Let us consider the remaining densities of Bethe roots. As far as $\gamma ^{(1)} (g,L), \ldots ,
\gamma ^{(4)} (g,L)$ are concerned, we only need to know $\sigma_0^{(0)}$, since $\sigma_{0}^{(1)}$ is zero.
Using results of \cite{FIR,FGR3}, we get
\ba
g \rightarrow \infty \Rightarrow \sigma_0^{(0)} = -4 \ln g + O(g^0) \, .
\ea
Hence, taking into account the previous equations, the leading strong coupling limit of  (\ref {gamma2}-\ref {gamma4}) turns out to be
\ba
\gamma^{(2)} (g,L) & = & \frac{\pi^2}{24\, m(g)}[(L-2)^3-(L-2)]+O\left ( e^{-\frac{\pi g}{\sqrt 2}}\right )\, ,
\label {gamma2strcoup}\\
\gamma^{(3)} (g,L) & = & -\frac{\pi}{3 \, m^2(g)} \ln g \, [(L-2)^3-(L-2)]+ O\left ( g^{-\frac {1}{2}}e^{\frac{2\pi g}{\sqrt 2}}\right ) \, , \label {gamma3strcoup} \\
\gamma^{(4)} (g,L) & = & 2
\frac{\ln ^2 g}{ m^3(g)} [(L-2)^3-(L-2)]
+ O\left ( g^{-\frac {3}{4}} \ln g \ e^{\frac{3\pi g}{\sqrt 2}}\right )\, .
\label{gamma4strcoup}
\ea

\section{Summary and conclusion}
\setcounter{equation}{0}

In this letter we have studied the sub-logarithmic terms $\gamma^{(n)}(g,L)$, $n\geq 1$, appearing in the high spin expansion (\ref {subl}) for the minimal anomalous dimension of twist operators in the $sl(2)$ sector of ${\cal N}=4$ SYM. We have found the
general expression (\ref {gammamain}), after solving via a recursive procedure the linear systems (\ref {Ssystem}): in particular, we have proved that $\gamma^{(1)}(g,L)=0$ and that $\gamma^{(n)}(g,2)= \gamma^{(n)}(g,3)=0$, $\forall n$. Then, we have found the connection (\ref {gammagen}) between $\gamma^{(n)}(g,L)$ and the generalised scaling functions appearing in the limit (\ref {jlimit}).
Finally, weak and strong coupling limits for $\gamma^{(n)}(g,L)$, with $n=2,\ldots 4$, are provided in Appendix A and in (\ref {gamma2strcoup} - \ref {gamma4strcoup}), respectively. In the large coupling regime and at leading order we would like to remark the appearance in $\gamma^{(n)}(g,L)$ of the mass gap for the O(6) Non-Linear Sigma Model: it would be nice to discover if a motivation for this result could be found on the string theory side of the correspondence, somehow miming \cite {AM}.

As for the future, the connection between $\gamma^{(n)}(g,L)$ and the generalised scaling functions describing the limit (\ref {jlimit}) is worth to be investigated, especially at strong coupling.
Another possible development is the study of the entire $sl(2)$ spectrum of anomalous dimensions \cite {BKP}, in the spirit of comparison to the recent developments on spiky strings \cite {FKT}.

\vspace {0.8cm}

{\bf Acknowledgements} We acknowledge the INFN grant {\it Iniziative specifiche FI11} and {\it PI14}, the international agreement INFN-MEC-2008 and the italian University PRIN 2007JHLPEZ "Fisica Statistica dei Sistemi Fortemente Correlati all'Equilibrio e Fuori Equilibrio: Risultati Esatti e Metodi di Teoria dei Campi" for travel financial support. The work of P.G. is partially supported by MEC-FEDER (grant FPA 2008-01838), by the Spanish Consolider-Ingenio 2010 Programme CPAN (CSD2007-00042) and by Xunta de Galicia (Conselleria de Educacion and grant PGIDIT06PXIB296182PR).

\appendix
\section{Weak coupling limit}
\setcounter{equation}{0}
\small{
The weak-coupling expansions for $\gamma^{(n)}(g,L)$, $n=2,3,4$ are given in the present appendix up to the order $g^{8}$. Such expansions can be easily obtained from the linear systems (\ref{Ssystem}) using a program of symbolic manipulation, in this case {\it Mathematica}$^{\textrm{\textregistered}}$. With little computational effort we were able to reach order $g^{22}$: for obvious reasons we do not give such higher orders here, but they are available
in the web page http://www-fp.usc.es/~grinza/gamma/gamma.html
\ba
&&\gamma^{(2)} (g,L)   = [(L - 2)^3 - (L - 2)]  \Bigg[
\frac {7} {24}\pi ^2 \zeta (3)  g^2+
\left (\frac {35} {144} \pi ^4 \zeta (3) - \frac {31}
{8} \pi ^2 \zeta (5) \right) g^4 + \nonumber \\
&&+\left (-\frac {73 \pi ^6 \zeta (3)} {4320} - \frac{155} {48} \pi ^4 \zeta (5) + \frac {635} {16} \pi ^2 \zeta (7) \right)
g^6+ \nonumber \\
&&+
 \left (\frac {7 \pi ^8 \zeta (3)} {1728} + \frac {91}{24} \pi ^2 \zeta (3)^3 + \frac {7} {60} \pi ^6 \zeta (5) + \frac{3175} {96} \pi ^4 \zeta (7) - \frac {17885} {48} \pi ^2 \zeta (9) \right)
g^8+ \ldots \Bigg]
\ea
\ba
&& \gamma^{(3)} (g,L)  = -[(L - 2)^3 - (L - 2)]  \Bigg[
\frac {7} {12} \pi ^2 (\ln 2 \, L + \gamma_E ) \zeta (3) \,g^2+
\nonumber \\
&& +
\frac {\pi ^2} {72}\Big(-3 (49 \zeta (3)^2 +
      186 \ln 2 \, \zeta (5) ) L + \gamma_E (35 \pi ^2 \zeta (3) -
      558 \zeta (5) +
   7 (11 L - 6) \pi ^2 \ln 2 \, \zeta (3)\Big) g^4 +
\nonumber \\
&& +\frac { \pi ^2 } {2160} \Big(135 (651 \zeta (3) \zeta (5) +
        1270 \ln 2 \, \zeta (7)) L -
     15 \pi ^2  (2046 \ln 2 \, \zeta (5) L + 385 \zeta (3)^2 L +
\nonumber \\
&& -
        1116 \ln 2 \, \zeta (5)) + \gamma_E  (-73 \pi ^4 \zeta (3) -
        13950 \pi ^2 \zeta (5) +
        171450 \zeta (7)) + (767 L -
         840) \pi ^4 \ln 2 \, \zeta (3) \Big) g^6+
\nonumber \\
&& -
\frac {\pi ^2 } {4320} \Big(\pi ^6 \ln 2 \, \zeta (3) (307 L -
        342) + 18 \pi ^4  (1184 \ln 2 \, \zeta (5) L +
        91 \zeta (3)^2 L - 1240 \ln 2 \, \zeta (5)+
\nonumber \\
&&  +
        105 \zeta (3)^2) -
     15 \pi ^2 (41910 \ln 2 \, \zeta (7) L +
        15011 \zeta (3) \zeta (5) L - 22860 \ln 2 \, \zeta (7) +
        1302 \zeta (3) \zeta (5))+
\nonumber \\
&&  -
     90  (L  (756 \ln 2 \, \zeta (3)^3 - 8649 \zeta (5)^2 -
           17780 \zeta (3) \zeta (7) -
           35770 \ln 2 \, \zeta (9) ) -
        392 \ln 2 \, \zeta (3)^3) +
\nonumber \\
&& - \gamma_E   (35 \pi ^6 \zeta (3) +
        1008 \pi ^4 \zeta (5) + 285750 \pi ^2 \zeta (7) +
        1260  (26 \zeta (3)^3 - 2555 \zeta (9) ) )
\Big) g^8 +
\ldots \Bigg]
\ea
\ba
&&\gamma^{(4)} (g,L)   =[(L-2)^3 - (L-2)]
\Bigg[ \frac {\pi ^2}{1920} \Big(3  (560 \ln^2 2 \, \zeta (3) -
        31 \pi ^2 \zeta (5))L^2 + 372 \pi ^2 \zeta (5) L  +
\nonumber \\
&&+
     3360 \gamma_E  \ln 2 \, \zeta (3) L - 155 \pi ^2 \zeta (5) +
     1680 \gamma_E ^2 \zeta (3) \Big) g^2 +
\nonumber \\
&&-\frac {\pi ^2}{11520} \Big(1440 \ln 2 \, (49 \zeta (3)^2 +
        93 \ln 2 \, \zeta (5)) L^2 -
     480 \gamma_E  (7 (11 L - 6) \pi ^2 \ln 2 \, \zeta (3) -
        3 L (49 \zeta (3)^2 +
\nonumber \\
&&+
           186 \ln 2 \, \zeta (5) ) ) -
     15 \pi ^2  ( (1904 \ln^2 2 \, \zeta (3) +
           1143 \zeta (7) ) L^2 +
\nonumber \\
&&-
        12  (112 \ln^2 2 \, \zeta (3) +
           381 \zeta (7) ) L + 1905 \zeta (7) ) +
     341 (3 L^2 - 12 L + 5 ) \pi ^4 \zeta (5) +
\nonumber \\
&&-
     240 \gamma_E ^2  (35 \pi ^2 \zeta (3) -
        558 \zeta (5) ) +
     14  (3 L^2 - 12 L + 5 ) \pi ^6 \zeta (3) \Big)g^4+
\nonumber \\
&&
-\frac { \pi ^2} {345600} \Big(-10800  (343 \zeta (3)^3 +
        3906 \ln 2 \, \zeta (3) \zeta (5) +
        3810 \ln^2 2 \, \zeta (7) ) L^2 -
     15 \pi ^4  ( (45872 \ln^2 2 \, \zeta (3)+
\nonumber \\
&& +
           62865 \zeta (7) ) L^2 -
        60  (1120 \ln^2 2 \, \zeta (3) +
           4191 \zeta (7) ) L +
        15  (1344 \ln^2 2 \, \zeta (3) +
           6985 \zeta (7) ) ) +
\nonumber \\
&& - 480 \gamma_E   (135 (651 \zeta (3) \zeta (5)+
           1270 \ln 2 \, \zeta (7)) L -
        15 \pi ^2  (2046 \ln 2 \, \zeta (5) L +
           385 \zeta (3)^2 L -
           1116 \ln 2 \, \zeta (5) ) +
\nonumber \\
&&+  (767 L -
            840) \pi ^4 \ln 2 \, \zeta (3) ) +
     900 \pi ^2  ( (5432 \ln 2 \, \zeta (3)^2 +
           12648 \ln^2 2 \, \zeta (5) +
           10731 \zeta (9) ) L^2 +
\nonumber \\
&& -
        12  (196 \ln 2 \, \zeta (3)^2+
           744 \ln^2 2 \, \zeta (5) + 3577 \zeta (9) ) L +
        17885 \zeta (9) ) +
     240 \gamma_E ^2  (73 \pi ^4 \zeta (3) +
        13950 \pi ^2 \zeta (5)+
\nonumber \\
&& - 171450 \zeta (7) ) -
     589  (3 L^2 - 12 L + 5 ) \pi ^6 \zeta (5) +
     40  (3 L^2 - 12 L + 5 ) \pi ^8 \zeta (3) \Big) g^6 +\nonumber
\ea
\ba
&& +\frac{ \pi ^2} {4838400} \Big(9277 \pi ^8 \zeta (5)  (3 L^2 -
        12 L + 5 ) +
     378 \pi ^{10} \zeta (3)  (3 L^2 - 12 L + 5 ) +
     210 \pi ^6  ( (4888 \ln^2 2 \, \zeta (3) +
\nonumber \\
&&+
           19431 \zeta (7) ) L^2 -
        612  (24 \ln^2 2 \, \zeta (3) +
           127 \zeta (7) ) L +
        15  (672 \ln^2 2 \, \zeta (3) +
           2159 \zeta (7) ) ) +
\nonumber \\
&&+
     302400 L  (L  (-4557 \zeta (3)^2 \zeta (5) - \ln 2 \, \
 (8649 \zeta (5)^2 + 17780 \zeta (3) \zeta (7) ) +
           7 \ln^2 2 \,  (82 \zeta (3)^3 -
              2555 \zeta (9) ) ) +
\nonumber \\
&&-
        392 \ln^2 2 \, \zeta (3)^3 ) -
     420 \pi ^4  (21  (4464 \ln 2 \, \zeta (3)^2 +
           14688 \ln^2 2 \, \zeta (5) +
           28105 \zeta (9) ) L^2 +
\nonumber \\
&&-
        60  (1092 \ln 2 \, \zeta (3)^2 +
           7440 \ln^2 2 \, \zeta (5) + 39347 \zeta (9) ) L +
        983675 \zeta (9)+
\nonumber \\
&& + 133920 \ln^2 2 \, \zeta (5) ) -
     3360 \gamma_E   (\pi ^6 \ln 2 \, \zeta (3) (307 L - 342) +
        18 \pi ^4  (1184 \ln 2 \, \zeta (5) L +
\nonumber \\
&&+
           91 \zeta (3)^2 L - 1240 \ln 2 \, \zeta (5) +
           105 \zeta (3)^2 ) -
        15 \pi ^2 (41910 \ln 2 \, \zeta (7) L +
           15011 \zeta (3) \zeta (5) L+
\nonumber \\
&& - 22860 \ln 2 \, \zeta (7) +
           1302 \zeta (3) \zeta (5)) -
        90  (L  (756 \ln 2 \, \zeta (3)^3 -
              8649 \zeta (5)^2 - 17780 \zeta (3) \zeta (7)+
\nonumber \\
&& -
              35770 \ln 2 \, \zeta (9) ) -
           392 \ln 2 \, \zeta (3)^3 ) ) +
     3150 \pi ^2  ( (29400 \zeta (3)^3 +
           427664 \ln 2 \, \zeta (3) \zeta (5) +
\nonumber \\
&&-
           4092 \zeta (3)^2 \zeta (5) +
           518160 \ln^2 2 \, \zeta (7) +
           644805 \zeta (11) ) L^2 -
        12  (13888 \ln 2 \, \zeta (3) \zeta (5) +
\nonumber \\
&&-
           1364 \zeta (3)^2 \zeta (5) + 30480 \ln^2 2 \, \zeta (7) +
           214935 \zeta (11) ) L + 1074675 \zeta (11) -
        6820 \zeta (3)^2 \zeta (5) ) +
\nonumber \\
&&+
     1680 \gamma_E ^2  (35 \pi ^6 \zeta (3) +
        1008 \pi ^4 \zeta (5) + 285750 \pi ^2 \zeta (7) +
        1260  (26 \zeta (3)^3 - 2555 \zeta (9) ) \Big)g^{8} +
\dots \Bigg]
\ea

\section{Computations of nonlinear terms}
\setcounter{equation}{0}

In this appendix we report on the computations of the generalised scaling functions
$f_n^{(r)}(g)$, with $1\leq r \leq 4$ and $0\leq n \leq 4-r$.
In order to do that, we have to
start from equation (\ref {Skeq2}). Forcing terms which contribute to those generalised scaling functions
are the ones containing  $\sum \limits _{h=1}^{L-2} \left[ \cos {t u_h} -1 \right]$.
We evaluate such an expression in the limit (\ref {jlimit})
by using results contained in Appendix A of \cite {FIR}. Using formul{\ae}
(A.3) and (A.5) of \cite {FIR}, we first have that
\ba
\sum _{h=1}^{L-2} \left[ \cos {t u_h} -1 \right]&=&-\int _{-c}^{c} \frac {dv}{2\pi}(\cos tv -1)
\sigma (v) - \frac {\pi}{6} \frac {t \sin tc}{\sigma (c)} - \nonumber \\
&-&  \frac {7\pi ^3}{360} \frac {t ^3 \sigma (c) \, \sin tc +3t^2 \sigma _1(c)\, \cos tc -3t \frac
{(\sigma _1(c))^2}{\sigma (c)}\sin tc +t \sigma _2 (c) \, \sin tc }{(\sigma (c))^4}
+ O\left ( \frac {j^n}{(\ln s )^5} \right ) = \nonumber \\
&=& -2 \int _{-\infty}^{+\infty} \frac {dk}{4\pi ^2} \hat \sigma (k) \left [ \frac {\sin (t+k)c}{t+k}
- \frac {\sin kc}{k} \right ]- \frac {\pi}{6} \frac {t \sin tc}{\sigma (c)} - \nonumber \\
&-& \frac {7\pi ^3}{360} \frac {t ^3 \sigma (c) \, \sin tc +3t^2 \sigma _1(c)\, \cos tc +t \sigma _2 (c) \, \sin tc }
{(\sigma (c))^4} + O\left ( \frac {j^3}{(\ln s )^3} \right )
\, ,
\ea
where $\sigma _m(c)$ denotes the $m$-th derivative of the density $\sigma (v)$ in $v=c$.
Restricting to the cases of our interest:
\ba
\sum _{h=1}^{L-2} \left[ \cos {t u_h} -1 \right]\Bigl |_{\frac {j^0}{(\ln s)^r}}&=& 0 \, , \quad r=1,2,3,4 \, ,
\nonumber \\
\sum _{h=1}^{L-2} \left[ \cos {t u_h} -1 \right]\Bigl |_{\frac {j}{\ln s}}&=& \frac {\pi ^2}{6} \frac {t^2}{(\sigma ^{(-1,0)})^2} \, , \quad
\sum _{h=1}^{L-2} \left[ \cos {t u_h} -1 \right] \Bigl |_{\frac {j^2}{\ln s}}= -\frac {\pi ^2}{3} t^2 \frac {\sigma ^{(-1,1)}}{(\sigma ^{(-1,0)})^3} \, , \nonumber \\
\sum _{h=1}^{L-2} \left[ \cos {t u_h} -1 \right]\Bigl |_{\frac {j^3}{\ln s}}&=& - \frac {\pi ^2}{2} t^2
\frac {(\sigma ^{(0,0)})^2}{(\sigma ^{(-1,0)})^4} -
\frac {\pi ^2}{6} \frac {t^2}{(\sigma ^{(-1,0)})^4}\left [ \frac {2}{3} \pi ^2 \frac {\sigma _2^{(-1,0)}}{\sigma ^{(-1,0)}}-3  (\sigma ^{(-1,1)})^2 + \frac {t^2}{6} \pi ^2 \right ] \, , \nonumber \\
\sum _{h=1}^{L-2} \left[ \cos {t u_h} -1 \right]\Bigl |_{\frac {j}{(\ln s)^2}}&=& -\frac {\pi ^2}{3} t^2\frac {\sigma ^{(0,0)}}{(\sigma ^{(-1,0)})^3} \, , \quad
\sum _{h=1}^{L-2} \left[ \cos {t u_h} -1 \right]\Bigl |_{\frac {j^2}{(\ln s)^2}}= \pi ^2 t^2 \frac {\sigma ^{(0,0)}\sigma ^{(-1,1)}}{(\sigma ^{(-1,0)})^4} \, , \nonumber \\
\sum _{h=1}^{L-2} \left[ \cos {t u_h} -1 \right]\Bigl |_{\frac {j}{(\ln s)^3}}&=& \frac {7 \pi ^4}{360}t^4 \frac {1}
{(\sigma ^{(-1,0)})^4}+ \frac {7 \pi ^4}{90}t^2 \frac {\sigma _2^{(-1,0)}}
{(\sigma ^{(-1,0)})^5}+ \frac {\pi ^2}{2} t^2 \frac {(\sigma ^{(0,0)} )^2}{(\sigma ^{(-1,0)})^4} \, , \nonumber
\ea
where we used notations of \cite {FIR}. After writing the systems for the Neumann modes we eventually realise that
\ba
f_0^{(r)} &=& 0 \, , \quad r=1,2,3,4 \, ; \quad
\frac {f_1^{(1)}} {{\sqrt {2}} g }=- \frac {\pi ^2}{3} \frac {\tilde S_1^{(1)}(g)}{(\sigma ^{(-1,0)})^2} \, , \quad
\frac {f_2^{(1)}} {{\sqrt {2}} g }= \frac {2\pi ^3}{3} \frac {\sigma ^{(-1,1)}}{(\sigma ^{(-1,0)})^3}\tilde S_1^{(1)}(g) \, , \nonumber \\
\frac {f_3^{(1)}} {{\sqrt {2}} g } &=& \frac {\pi ^3}{(\sigma ^{(-1,0)})^4}\left [ \left ( \frac {2}{9}\pi ^2
\frac {\sigma _2^{(-1,0)}}{\sigma ^{(-1,0)}} - (\sigma ^{(-1,1)})^2
+ (\sigma ^{(0,0)})^2 \right ) \tilde S_1^{(1)}(g) + \frac {\pi ^2}{18}\tilde S_1^{(2)}(g) \right ] \, ;\label {f13} \\
\frac {f_1^{(2)}} {{\sqrt {2}} g } &=& \frac {2\pi ^3}{3} \frac {\sigma ^{(0,0)}}{(\sigma ^{(-1,0)})^3}\tilde S_1^{(1)}(g) \, , \quad
\frac {f_2^{(2)} }{{\sqrt {2}} g } =- 2\pi ^3  \frac {\sigma ^{(0,0)} \sigma ^{(-1,1)}}{(\sigma ^{(-1,0)})^4}\tilde S_1^{(1)}(g) \, ; \nonumber \\
\frac {f_1^{(3)}} {{\sqrt {2}} g } &=&-  \frac {\pi ^3} {(\sigma ^{(-1,0)})^4}\left [ \frac {7 \pi ^2}{180} \tilde S_1^{(2)}(g) + \left ( \frac {7 \pi ^2}{45}  \frac {\sigma _2^{(-1,0)}}
{\sigma ^{(-1,0)}}+   (\sigma ^{(0,0)} )^2  \right ) \tilde S_1^{(1)}(g) \right ]
\, .\nonumber
\ea

\end{document}